\magnification=1200
\baselineskip=0.55 truecm
\hsize 16 truecm \vsize 22.5 truecm
\hoffset=+0.3truecm 
\tolerance=10000
\def\pni{\par \noindent}
\def\vsh{\vskip 0.25truecm}
\def\vs{\vskip 0.5truecm}
\def\vvs{\vskip 1.0truecm}
\def\vvvs{\vskip 1.5truecm}
\def\vsp{\vsh \par}
\def\vsn{\vsh \pni}
\def\q{\quad}
\def\qq{\qquad}
\def\cen{\centerline}
\def\eg{{\it e.g.}\ }
\def\ie{{\it i.e.}\ }
\def\bar{\overline}
\def\dis{\displaystyle}
\def\ds{\displaystyle}

\def\l{\left} \def\r{\right}
\def \rec#1{{1\over{#1}}}

\def\rf#1{\item{$[{#1}]$ \ }}

\def\NN{{\rm I\hskip-2pt N}}
\def\RR{\vbox {\hbox to 8.9pt {I\hskip-2.1pt R\hfil}}}
\def\II{{\rm I\hskip-2pt I}}
\def\CC{{\rm C\hskip-4.8pt \vrule height 6pt width 12000sp\hskip 5pt}}
\font\ten=cmr10

\font\title=cmbx10 scaled \magstep1
\pageno=140
\ten
\cen{EXTRACTA MATHEMATICAE \ Vol. {\bf 10}, N\'um. 1, 140---154 (1996)}
\vsh\hrule
\vskip 1.1truecm
\cen{\title The Fractional Langevin Equation:}
\vsh
\cen{\title  Brownian Motion Revisited}
\vs
\ten
\cen{Francesco MAINARDI and Paolo PIRONI}
\vsh
\cen{{\it Department of Physics, University of Bologna, Via Irnerio 46,}}
\cen{{\it I-40126 Bologna, Italy; e-mail: mainardi@bo.infn.it}}
\cen{{\it www.fracalmo.org}}
\vs
\line{AMS {\it Subject Class} (1991): 60J60, 60J65, 45E10, 45J05,
   44A10, 33B20. \hfill}
\vvs
\cen{{INTRODUCTION}}\vsp
It is well known that the concept of  diffusion is associated
with random motion of particles in space, usually denoted as
Brownian motion, see \eg [1-3].
Diffusion is considered normal	when
 the mean squared displacement of the particle during a time interval
becomes, for sufficiently long intervals,
a linear function of it. When this linearity breaks down,
degenerating in a power law with exponent $\alpha >0 $ different from one,
diffusion is referred to be anomalous: slow if $ 0<\alpha <1\,,$
fast if $\alpha >1\,; $ see \eg [4].
\vsp
According to the classical approach started by Langevin
 and known as the Einstein-Ornstein-Uhlenbeck theory of
Brownian motion,
normal diffusion  and Brownian motion
are associated with Langevin  equation.
More specifically,  
the classical Langevin equation 
addresses the dynamics of a Brownian particle
through Newton's law    by incorporating the effect of the Stokes
fluid friction and  that  of thermal
fluctuations in the vicinity of the particle into a random
force with suitably assigned properties. These properties are derived
from the requirement that the particle velocity
asymptotically attains a stationary Maxwellian distribution.
Over the period of the diffusing particle, the random force
arising from molecular collisions undergoes such a rapid fluctuations that
it is approximated well by a white noise.
For large time intervals $t$, it emerges that the mean squared
displacement 
becomes proportional to $t$
with the diffusion coefficient being a half of the
proportionality constant, in the one  dimensional case.
\vsp
\hrule
\vsn
The present e-print is a reproduction of the  contribution published in 1996,
 so it  represents our knowledge of that early time. Since 1996  many papers have appeared on the topic
 in view of the rapidly developing theory of fractional diffusion processes.
 The corresponding author (FM) intends to submit an up-dated review on the topic, so he is grateful to arXiv readers 
 for any comment and suggestion they may have on this  e-print.    
\vfill\eject
\vsp
In Sect. 1. we	summarize the salient mathematical aspects of the classical
Langevin equation, showing the exponential time decay
of the velocity correlation function and the linear long-time behaviour
of the mean squared displacement.   We also consider the generalized
version of the Langevin equation introduced by Kubo [2] to
account  for a general retarded effect of the frictional force
and the two fluctuation-dissipation theorems by Kubo.
\vsp
In Sect. 2, following the approach originally
started by Widom [5] and Case [6]  and resumed in [3-4],
 we shall consider the modification of the Langevin
equation on the basis of hydrodynamics, which takes
into account the  added mass  and the Basset-Boussinesq  retarding
force.
We shall improve the analysis of the previous authors,
interpreting the  retarding force in the framework of {\it Fractional
Calculus} and  providing the analytical expressions of the
autocorrelation functions (both for velocity and  random force)
and of the  mean squared displacement.
\vsp
We shall conclude noting that,
for  "not heavy" Brownian particles,
there is the possibility for anomalous diffusion,
with $\alpha > 1 \,,$ in a long time interval, before the normal
diffusion is established.
\vvvs
\cen{{1. THE CLASSICAL AND GENERALIZED LANGEVIN EQUATIONS}}\vs\vsp
According to the classical Langevin approach  the  dynamics in one
dimension for a Brownian particle is described by
$$ \eqalignno{ {dX \over dt} &= V \,,	&(1.1)\cr
m \, {dV\over dt} &= F\,,  &(1.2)\cr} $$
 where $m$ is the particle mass,  $X= X(t)$,  $V= V(t)$  are
the particle position and  velocity, and $F$ is
the force acting on the particle from
molecules of the fluid surrounding the Brownian particle.
\vsp
The force $F$ may be divided into two parts.
The first part is the {\it frictional force} and is taken
to be proportional to the particle velocity \ie
$$ F_v = - {m\over \sigma }\, V \,,\eqno(1.3)$$
 where	$1/\sigma $ is the friction coefficient for unit mass.
One usually introduces	the {\it mobility coefficient} as
$$ \mu	:= {\sigma  / m}\,. \eqno(1.4)$$
If the Stokes law is assumed for a spherical particle of radius $a\,,$
see \eg [7],  we have
$$ 1/\mu =  6 \pi \, a \,\rho_f \,\nu\,,  \eqno(1.5)$$
where $\rho _f$ and $\nu $ are the density and the kinematic viscosity
of the fluid, respectively.
Introducing  the characteristic parameters
$$ \tau _0 := {a^2 / \nu }\,, \q \chi	:= {\rho _p / \rho _f}\,,
  \eqno(1.6)$$
where $\rho _p$ is the particle density, we obtain
$$ \rec{\mu } = {9 \over 2\chi	}\, {m\over \tau_0 }
   \, \Longrightarrow \,
\sigma := {2\over 9}\, \chi   \, \tau_0 \,. \eqno(1.7)$$
\vsp
The second part of the force, arising from rapid thermal
fluctuations, is regarded as random, independent of the motion of the
particle. This part is called  the {\it random force} and is hereafter
denoted by $R(t)$.
\vsp
Then (1.2) is written as a stochastic equation as
  $$		     {dV\over dt} =
 -\rec{\sigma }\,  V(t) +  \rec{m} \, R(t)\,,\eqno (1.8)$$
and it is referred to as the  {\it classical Langevin equation}.
\vsp
It is assumed that the stochastic processes $V(t)$ and $R(t)$
be {\it stationary}. 
This means that the respective
{\it autocorrelation functions} $C_V$ and $C_R$,
$$ C_V(t_0,t) :=\langle \,V(t_0)\, V(t_0 +t)\, \rangle = C_V(t)\,,
   \q t>0\,,
   \eqno(1.9)$$
and
$$ C_R(t_0,t):=  \langle \,R(t_0)\, R(t_0 +t)\, \rangle = C_R(t)\,,
  \q t>0\,,
  \eqno(1.10)$$
do not depend on $t_0\,. $ Hereafter  we will assume $t_0 =0\,. $
\vsp
As a consequence, because of the  Wiener-Khintchine theorem [3],
the   {\it power spectra} or
{\it power spectral densities} $I_V(\omega)$ and $I_R(\omega )$
($\omega \in \RR$)
are provided by the Fourier transforms of the  respective
autocorrelation functions. We write
$$ I_V(\omega ) = \widehat C_V(\omega) :=
 \int_{-\infty}^{+\infty}\!\! C_V(t)\, {\rm e}^{\dis -i\omega\,t}\,dt  \,, \q
   C_V(t) = \rec{2\pi}\,\int_{-\infty}^{+\infty}  \!\! I_V(\omega)\,
   {\rm e}^{\dis +i\omega\,t}\,d\omega \,, \eqno(1.11)$$
and
$$ I_R(\omega ) = \widehat  C_R(\omega) :=
 \int_{-\infty}^{+\infty}\!\! C_R(t)\, {\rm e}^{\dis -i\omega\,t}\,dt  \,, \q
   C_R(t) = \rec{2\pi}\,\int_{-\infty}^{+\infty}  \!\! I_R(\omega)\,
   {\rm e}^{\dis +i\omega\,t}\,d\omega \,. \eqno(1.12)$$
\vsp
We assume that any process $f(t)$ be causal, \ie vanishing for $t<0\,,$
so that the Fourier transform $\widehat  f(\omega )$ of $f(t)$
is   related to  the Laplace transform	   by the identity
$$ \widehat  f(\omega) = \bar f(s) \vert_{s=i\omega}\,, \q
   \bar f(s) := \int_0^{\infty} \! {\rm e}^{-st}\, f(t)\, dt \,,
     \q s \in \CC\,.  \eqno(1.13)$$
\vsp
It is assumed that the random force has zero mean and is uncorrelated
to  the  particle velocity at initial time $t=0\,; $
in other words,
$$ \langle \, R(t)\, \rangle= 0\,, \q
  \langle \, V(0) \, R(t) \, \rangle = 0\,, \q t>0\,.  \eqno(1.14)$$
Furthermore, if the Brownian particle has been kept for a
sufficiently long time in the fluid at (absolute) temperature $T$,
the equipartition law
 $$ m \, \langle \, V^2(0)\, \rangle  =  k \, T\,, \eqno(1.15)$$
where	  $k$ is the Boltzmann constant,
is assumed for the energy distribution.
 \vsp
It can be shown (see below)
that the previous assumptions  lead to the
following relevant results
$$
   C_V(t) =  \langle \, V^2(0)\, \rangle  \, {\rm e}^{\dis -t/\sigma }
   =  {k\, T\over m}   \, {\rm e}^{\dis -t/\sigma }
\,,\eqno(1.16)$$
and
$$ C_R(t) = {m^2 \over \sigma } \,
  {\langle \, V^2(0)\, \rangle} \, \delta (t) =
	  {m\, k\, T\over \sigma } \, \delta (t)\,.
\eqno(1.17)$$
\vsp
The result (1.16)  shows that the velocity autocorrelation
function  decays exponentially in time with the decay constant
$\sigma  \,,$ while (1.17) means that
the power spectrum of $R(t)$ is  to be {\it white},
\ie independent on frequency, resulting
$$ I_R(\omega ) \equiv I_R =
     {m\, k\, T\over \sigma } \,.   \eqno(1.18)$$
\vsp
The two results can be generalized for the  so-called
{\it generalized Langevin equation} introduced by Kubo [2],
$$		   {dV\over dt} =
 -\int_0^t \gamma(t-\tau) \,  V(\tau ) \,d\tau
+  \rec{m} \, R(t)\,,\eqno (1.19)$$
where the function $\gamma (t)$ represents a retarded effect of the
frictional force. For this case Kubo introduced two
{\it fluctuation-dissipation theorems} that, using the
Laplace transforms, read   respectively
$$ \bar C_V(s) =
   { \langle \, V^2(0)\, \rangle \over
      s + \bar \gamma (s)}\,, \eqno(1.20)$$
and
$$
 \bar C_R(s) =	m^2 \,	 { \langle \, V^2(0)\, \rangle}\,
     \bar \gamma (s)\,. \eqno(1.21)$$
\vsp
From the comparison between (1.8) and (1.19),
we recognize that the classical case can be obtained from the generalized
one interpreting the convolution in (1.19) in the generalized sense (see
\eg [8]) and putting
$$   \gamma (t) = \rec{\sigma } \, \delta (t)\,
	\Longleftrightarrow \,	\bar \gamma (s)  = \rec{\sigma } \,,
\eqno(1.22)$$
where $\delta (t)$ denotes the delta Dirac distribution.
\vsp
In Appendix A we prove the statements (1.20) and (1.21), which
thus reduce to the classical results (1.16) and (1.17)
accounting for (1.22).
\vsp
It can be readily shown that the mean squared displacement of a particle,
starting at the origin at $t_0 =0\,,$ is given by
$$ \langle\,X^2(t)\, \rangle=2\,  \int_0^t \!\! (t-\tau)\,
C_V(\tau)\,d\tau = 2\, \int_0^t d\tau_1 \int_0^{\tau _1}
  \!\! C_V(\tau )\, d\tau \,. \eqno(1.23)$$
For this it is sufficient  to recall that
 $X(t)=\int_{0}^t V(t')\,dt'\,,$ and to use the definition (1.9) of
$C_V(t)\,.$
\vsp
For the classical case, $C_V(t)$ is provided by (1.16), so that we
obtain from (1.23)
 $$ \langle\,X^2(t)\, \rangle =  2  \, \langle \, V^2(0)\, \rangle
	\,\sigma  \,
  \left[\,t- \sigma \, \left(1-{\rm e}^{\dis - t/\sigma }\right )
   \,\right]\,.\eqno(1.24) $$
Introducing  the {\it diffusion coefficient} $D$ as
$$ D = \sigma \, {\langle \, V^2(0)\, \rangle}
    =\mu  \, k \, T \,,
     \eqno(1.25)$$
where we have used  (1.4) and (1.15),
we can deduce  for large times	the well-known property
    $$	 \langle\,X^2(t)\, \rangle =
    2 D\, t \l\{1 +
   O\l[ \l({t/ \sigma}\r)^{-1} \r] \r\}
\,, \q {\rm as} \q t\to \infty \,. \eqno(1.26)$$
The relationship stated in  (1.25), which is called the
{\it Einstein relation}, provides us with a very good basis of
experimental verification that Brownian motion is in fact related to the
thermal motion of molecules.
We also note the following relevant results
for the {\it diffusion coefficient}
$$ D = \lim_{t \to \infty} \,
  {\langle\,X^2(t)\, \rangle \over 2\,	t} =
   \int _0^{\infty} \!\! C_V(t) \, dt = \bar C_V(0)
\,. \eqno(1.27)$$
\vvvs
\cen{{2. THE FRACTIONAL LANGEVIN EQUATION}}\vs\vsp
On the basis of hydrodynamics the equation of motion (1.8) is not
all correct since it ignores the effects of the added mass and of
the retarded viscous force, which are due to the acceleration of the
particle, see \eg [5-7].
 \vsp
The added mass effect introduces a modification in the L.H.S. of (1.2)
in that it requires to substitute the mass of the particle
with the so-called effective mass, namely
    $$ m \to m_e := m + \rec{2}\, m_f =
  m\, \l( 1 + \rec{2\chi  }\r)
     \,, \eqno(2.1)$$
where $\chi   = \rho _p/\rho _f$ according to (1.6).
As a consequence, in order to do not change the mobility coefficient
in the Stokes drag, we have to introduce $\sigma _e$ such that
$$ \mu := \sigma /m = \sigma _e/m_e \,, \eqno(2.2)$$
namely, recalling (1.6-7),
    $$
  \sigma _e:= \sigma \, \l(1 +\rec{2\chi  }\r)
    =	{2\chi	+1\over 9}\, \tau_0  \,,\eqno(2.3)$$
where $\tau_0 = a^2/\nu $.
\vsp
With respect to the classical analysis, it turns out that
the  added mass effect, if it were present alone,  would be  only
to lengthen the time scale ($\sigma  \to \sigma _e >\sigma \,$),
slowing down the exponential decay  for
the velocity correlation function (1.16) and for the mean
square displacement (1.24), but without modifying the
value of the diffusion coefficient. 
\vsp
The retarded viscous force  effect is due to an additional term to the
Stokes drag, which is related to the history of  the particle acceleration.
This additional drag force,
proposed independently by Boussinesq  [9] and Basset [10] in earlier
times, is nowadays referred to as the {\it Basset force}.
As a consequence, the
frictional force (1.3-5)  is to be substituted as follows
$$ F_v = -  6 \pi \, a \,\rho_f \,\nu\,
       \l\{ V(t) +
   {a \over \sqrt{\pi \nu }}   \,   \int_{-\infty}^t
{d V(\tau )/d\tau \over
 \sqrt{t-\tau}}\,d\tau	\r\}\,.\eqno(2.4)$$
Using (1.6-7) and requiring the {\it causality}
of the processes,
we can re-write (2.4) as
$$ F_v = - {9 \over 2\, \chi  } \,m\,  \l[
     \rec{\tau_0 }\,   V(t)  +	\rec{\sqrt{\tau_0 }} \,
       B(t) \r]\,, \eqno(2.5)$$
where
$$ B(t):=  \rec{\sqrt{\pi}}\,
     \int_{-\infty}^t {d V(\tau )/d\tau \over \sqrt{t-\tau}}\,d\tau
  =	\rec{\Gamma (1/2)} \,
     \int_{0^-}^t {d V(\tau )/d\tau \over \sqrt{t-\tau}}\,d\tau   \,.
\eqno(2.6)$$
The lower limit of the integral has been written as $0^-$
to account for the possible discontinuity in the velocity
particle at $t=0$.
Basing on the {\it Fractional Calculus} recalled in
Appendix B,   we can  write, see (B.11-12) and (B.6),
$$ B(t) =
    D_0^{1/2} \, V(t)
   = \Phi_{-1/2}(t) \,*\, V(t)\,, \eqno(2.7)$$
where
$ D_0^{1/2} $ denotes the fractional derivative of order $1/2$ and
 $$   \Phi_{-1/2}(t) :=  {\; t^{-3/2}\over \Gamma (-1/2) }
	   = - {\; t^{-3/2}\over 2\sqrt{\pi} } \,. \eqno(2.8)$$
\vsp
Then, adding the random force $R(t)$, the complete Langevin equation (1.2)
turns out to be
$$		  {dV\over dt} =
 -\rec{\sigma_e }\,\l[
  1 + \sqrt{\tau_0 } \,D_0^{1/2}\r] \, V(t)
+  \rec{m_e} \, R(t)\,.\eqno (2.9)$$
We agree to refer to (2.9) as to the {\it fractional Langevin equation}.
\vsp
We recognize that our fractional Langevin equation
is a particular case of
the generalized Langevin equation (1.19) with
$$ \gamma (t) = \rec{\sigma_e }\,
     \l[ \delta (t) - \sqrt{\tau_0 } \, {\; t^{-3/2} \over 2 \sqrt{\pi}}
     \r]
 \,  \Longleftrightarrow \, \bar \gamma(s) =
	 \rec{\sigma_e }\,
     \l[ 1 + \sqrt{\tau_0 } \, s^{1/2} \r] \,. \eqno(2.10) $$
Consequently, we can use (2.10)
to compute  the correlations functions	$C_V(t)\,, \, C_R(t)$
starting from their  Laplace transforms (1.20-21), respectively.
Then, the mean squared displacement
can be derived from $C_V(t)$ according to (1.23).
\vsp
Let us first consider the random force. The inversion of the
Laplace transform $\bar C_R(s)$ yields
$$ C_R(t) =
 m_e^2\,
  { \langle \, V^2(0)\, \rangle}  \, \gamma (t) \,, \eqno(2.11)$$
where $\gamma (t)$ is provided by (2.10).
We thus recognize that
for our fractional Langevin equation the random force cannot
be longer represented uniquely by a white noise;  an
additional "fractional" noise is  present due to the term $t^{-3/2}$
which, as formerly noted by
Case  [6], is to be interpreted in the generalized sense
of tempered distributions [8].
\vsp
Let us now consider  the velocity correlation.
Using (1.20) and (2.10) it turns out
$$ \bar C_V(s) =
    { {\langle \, V^2(0)\, \rangle} \over
      s +\l[1+ \sqrt{\tau_0 } \, s^{1/2}\r]/ \sigma_e  }
 = { {\langle \, V^2(0)\, \rangle} \over
      s + \sqrt{\beta /\sigma_e} \, s^{1/2} +1/\sigma_e  }
\,, \eqno(2.12)$$
where, because of (2.3) and (1.6),
$$ \beta :=
       {\tau_0 \over \sigma _e} =  {9\over 2\chi   +1} =
     {9 \rho _f \over 2 \rho _p + \rho _f}
\,. \eqno(2.13)$$
We note from (2.13) that $ 0 < \beta < 9$, the limiting
cases occurring  for $\chi  = \infty$  and $\chi   =0$,
 respectively.
We also recognize that the effect of the
Basset force is expected to be negligible for $\beta \to 0\,,$
\ie  for particles
which are  sufficiently
heavy with respect to the fluid ($\rho _p \gg \rho _f$).
\vsp
As far as we know, at least in this context, an explicit inversion
of the	Laplace transform (2.12) in terms of elementary functions
has not yet been carried out.
Widom [5] and  Case [6] have only provided  integral representations of
the velocity correlation function, from which they have derived
the long-time asymptotic behaviour ($ \propto t^{-3/2}$).  In our
notation, applying the asymptotic theorem for $s \to 0$  to (2.12), we get
as $t \to \infty$
$$ C_V(t) \simeq
    {\langle \, V^2(0)\, \rangle}\,
 {\sqrt{\beta } \over  2 \sqrt{ \pi}}
 \,   \l({t\over \sigma_e}\r)^{-3/2}
   = \,   {{\langle \, V^2(0)\, \rangle} \over 2 \beta \sqrt{\pi}}
 \, \l({t\over \tau_0 }\r)^{-3/2}  \,. \eqno(2.14)$$
The presence of such a long-time tail, pointed out also in [3-4],
was first observed by Alder and Wainwright [11] in a computer
simulation of velocity correlation functions.
\vsp
The explicit inversion of (2.12) is hereafter carried out, basing on our
previous analysis of the original and generalized
Basset problems,  in the
framework of  the {\it Fractional Calculus}
 and {\it Mittag-Leffler functions}
[12-14]. For this aim let
us recall the following Laplace transform pairs

\def\ss{{s}^{1/2}}   
$$
 {1\over (\ss -a_+)\,(\ss-a_-)} \,\div \,  \rec{a_+ - a_-} \,\l[
 a_+\,	E_{1/2} (a_+\,\sqrt{t}) - a_-\,  E_{1/2} (a_-\,\sqrt{t})\r]\,,
  \eqno(2.15)
   $$
$$
     {1\over  (\ss -a)^2}
\,\div \,    E_{1/2} (a\,\sqrt{t}) \, \l[ 1 +2\, a^2\,t \r]
       + 2\,a\, \sqrt{{t/ \pi}} \,, \eqno(2.16)
    $$
where
    $$ E_{1/2} (a\,\sqrt{t})  :=
   \sum_{n=0}^\infty  {a^n\, t^{n/2}\over \Gamma (n/2+1)}
     =	 {\rm e}^{\ds a^2\,t}\, {\rm erfc} (-a\,\sqrt{t}) \eqno(2.17) $$
denotes the {\it Mittag-Leffler  function} of order $1/2\,. $
In fact, re-writing (2.12) as
$$  \bar C_V(s) = { {\langle \, V^2(0)\, \rangle} \over
      (\ss -a_+)\,(\ss-a_-)  }	   \eqno(2.18)		    $$
where
$$     a_{\pm} =
     {-\sqrt{ \beta } \pm (\beta -4)^{1/2} \over  2\, \sqrt{\sigma _e}}
 \q {\rm if}\q	\beta  \ne 4 \,,
  \qq a_{\pm} =  a =-\rec{\sqrt{\sigma _e}}
  \q {\rm if}\q  \beta	= 4
   \,, \eqno(2.19)$$
we easily obtain the required $C_V(t)$	in terms of {\it Mittag-Leffler
functions}, as pointed out in (2.15-16).
\vsp
Furthermore,
it can be proved  that $C_V(t)$
results for $t>0$ a decreasing	function,
completely monotonic, \ie $(-1)^{n}\, C_V^{(n)} (t) > 0\,, $
with the asymptotic behaviour given by (2.14), for any
physical value of $\beta \,.$
\vsp
In order to compute
${\langle\,X^2(t)\, \rangle}$, according to (1.23)
we have to consider the 2-fold
primitives of the functions in the R.H.S. of (2.15-16), vanishing at $t=0$.
In particular, the repeated integral for the {\it Mittag-Leffler
function} turns out
 $$ \eqalign{ I_0^2 \, E_{1/2}(a\sqrt{t}) &=
      \sum_{n=0}^\infty  {a^n\, t^{n/2+2}\over \Gamma (n/2+3)} \cr
   &= \rec{a^4}\, \l[
     E_{1/2}(a\sqrt{t})  -1 - 2a \,{t^{1/2} \over \sqrt{\pi}}
  - a^2 \,t	- {4\over 3} a^3\, {t^{3/2} \over \sqrt{\pi}}
      \r] \,.\cr} \eqno(2.20)$$
The asymptotic behaviour of ${\langle\,X^2(t)\, \rangle}$
as $t\to \infty$ can be easier obtained from its Laplace
transform for $s\to 0\,, $ and reads
$$ {\langle\,X^2(t)\, \rangle}	= 2 D\, t \l\{1 +
   O\l[ \l({t/ \sigma _e}\r)^{-1/2} \r] \r\}
\,, \q {\rm as} \q t\to \infty \,, \eqno(2.21)$$
where
$$ D= \bar C_V(0) = \sigma _e \, {\langle\,V^2(0)\, \rangle}
    = \mu \, k\, T \,. \eqno(2.22)$$
Note that in the RHS of (2.22) we have used the energy equipartition law
(1.15) with the effective mass and (2.2).
\vsp
The explicit expressions of the velocity autocorrelation function
and of the displacement variance   are given in  [14-15].
\vvvs
\cen{{CONCLUSIONS}}\vs\vsp
In this paper we have revisited the Brownian motion
on the basis of the {\it fractional Langevin equation}	(2.9),
which turns out to be a particular case of  the generalized
Langevin equation (1.19) introduced by Kubo on 1966.
\vsp
The importance of our approach
is  to	model the  Brownian motion
more realistically  than the usual one based on the classical Langevin
equation (1.8), in that
it takes into account also the retarding effects due to
hydrodynamic backflow, \ie the added mass and the Basset memory drag,
as pointed out in (2.1) and (2.4), respectively.
\vsp
On the basis of the two {\it fluctuation-dissipation theorems}
(recalled in the Appendix A) and  of the techniques of
the {\it Fractional Calculus} (recalled in the Appendix B),
we have provided the analytical expressions
of the correlation functions (both for the random force and  the
particle velocity)  and of the mean squared particle displacement.
Consequently, the well-known results of the classical theory
of the Brownian motion	have been properly generalized.
\vsp
The  random force  has been shown to
be represented by a superposition of the usual white noise
with a "fractional" noise, as pointed out in   (2.10-11),
\vsp
The  velocity  correlation function $C_V(t)$
exhibits a different behaviour from the classical case: it is no longer
expressed by a simple exponential   but  by a combination of
Mittag-Leffler functions of order 1/2, according to (2.15-19).
 As a consequence, one can derive for $C_V(t)$	a
slower decay, proportional to $t^{-3/2}$ as $t \to \infty\,, $
which indeed is more realistic than the usual exponential one,
also in view of numerical simulations.
\vsp
Finally,  the mean squared  displacement
has been shown to maintain, for sufficiently long times,
the linear behaviour  which is typical of normal diffusion, with the same
diffusion coefficient	of the classical case, as seen in (2.21-22),
\ie  ${\langle\,X^2(t)\, \rangle} \simeq 2 \, D\, t \,. $
However, the Basset memory force,
which is responsible of the algebraic decay of the velocity correlation
function, induces  a retarding effect ($\propto t^{1/2}$) in the
establishing   of the linear behaviour, which is  relevant when
the parameter $\beta $ introduced in (2.13) is big enough.
From numerical computations this effect is seen to be
evident when $ 0< \rho _p < 2\, \rho _f$, \ie for "not heavy"
Brownian particles; in these cases one can get a best fit
in a long time interval with the law
    ${\langle\,X^2(t)\, \rangle} \simeq 2 \, D_*\, t^{\alpha} \,, $
with $0<D_*\, \sigma ^{\alpha -1}<D $ and $1<\alpha <2\,,$
which appears as a manifestation of fast anomalous diffusion [14-15].
\vvvs
\cen{{ACKNOWLEDGEMENTS}}\vs\vsp
We are	grateful to F. Tampieri for useful discussions.
This research was part\-ly sup\-ported by MURST (60\% grants)
and by INFN (sez. Bologna).
\vvvs
\cen{{APPENDIX A}}\vs\vsp
Let us consider the {\it generalized Langevin equation} (1.19),
that we write as
$$   R(t) = m \, \l[ \dot V(t) + \, \gamma(t) *   V(t)\r]
\,,\eqno (A.1)$$
where $\cdot \null $ denotes time differentiation and $*$   time
convolution.
 The assumption of stationarity for the stochastic processes
along with the following hypothesis
$$ \langle \, R(t)\, \rangle= 0\,, \q
   \langle \, V(0) \, R(t) \, \rangle = 0\,, \q t>0\,,	\eqno(A.2)$$
allows us to derive, by using the Laplace transforms,
the two {\it fluctuation-dissipation theorems}
$$ \bar C_V(s) := \bar{{\langle\,  V(0) \, V(t)\, \rangle}}
   =  { \langle \, V^2(0)\, \rangle \over s + \bar \gamma (s)}
     \,, \eqno(A.3)$$
and
$$
 \bar C_R(s) := \bar{{\langle\,  R(0) \, R(t)\, \rangle}}
=  m^2 \,   { \langle \, V^2(0)\, \rangle}\,
     \bar \gamma (s)\,. \eqno(A.4)$$
Our derivation is alternative to
the original one by Kubo
who used Fourier transforms [2]; furthermore, it  appears useful
for the treatment of our {\it fractional Langevin equation}.
\vsp
Multiplying both sides of (A.1) by $V(0)$ and averaging, we
obtain
$${\langle\,  V(0) \, \dot V(t)\, \rangle}
  + \, \gamma (t) *   {\langle\,  V(0) \, V(t)\, \rangle} =0\,.
    \eqno(A.5) $$
The application of the Laplace transform to both sides of (A.5) yields
$$ s\, \bar{{\langle\,	V(0) \, V(t)\, \rangle}} \, - \,
    {\langle\,	V^2(0)\, \rangle} \, + \,\bar \gamma(s) \,
      \bar{{\langle\,  V(0) \, V(t)\, \rangle}} =0\,, \eqno(A.6)$$
from which  we just obtain (A.3).
\vsp
Multiplying both sides of (A.1) by $R(0)$ and averaging, we
obtain
$$ C_R(t) :=
       {\langle\,  R(0) \, R(t)\, \rangle}
	    = m^2 \, \l[
    {\langle\, \dot  V(0) \, \dot V(t)\, \rangle}
  + \, \gamma (t) *  {\langle\, \dot V(0) \,  V(t)\, \rangle}\,\r]
\,. \eqno(A.7)$$
Noting that, by the stationary condition,
$$ {\langle\,  \dot V(0) \,  V(0)\, \rangle} =0 \,, \q
  {\langle\,  \dot V(0) \,  V(t)\, \rangle} =
 - \, {\langle\,  V(0) \,  \dot V(t)\, \rangle}\,, \eqno(A.8)$$
the application of  the Laplace transform to both  sides
of (A.7) yields
$$ \bar C_R(s) = m^2 \,
\l\{ s\, \bar {{\langle\, \dot V(0) \,	V(t)\, \rangle}}
	- \bar \gamma (s) \,
 \l[ s \, \bar {{\langle\,   V(0) \, V(t)\, \rangle}}-
     {\langle\,   V^2(0)\, \rangle}   \r]  \r\} \,. \eqno(A.9) $$
Since
 $$ \bar {{\langle\,  \dot V(0) \,  V(t)\, \rangle}}
 = - \, \bar {{\langle\,  V(0) \, \dot V(t)\, \rangle}}
 = -s \, \bar {{\langle\,  V(0) \,  V(t)\, \rangle}}
  + {\langle\,	V^2(0) \, \rangle}\,,  \eqno(A.10) $$
we get
$$ \bar C_R(s) = m^2 \,
\l\{ s\, \l[ -s\, \bar C_V(s)  +  {\langle\,  V^2(0) \, \rangle}
	- \bar \gamma (s) \, \bar C_V(s) \,\r]
 + \bar\gamma (s)\, {\langle\,	 V^2(0)\, \rangle}   \r\} \,,
   \eqno(A.11) $$
from which, accounting for (A.3), we just obtain (A.4).
\vfill\eject
\cen{{APPENDIX B}}\vs\vsp
Here we  recall
the essentials of Riemann-Liouville {\it Fractional Calculus}
basing on [16-20],
and we
interpret the Basset force
in terms of a fractional derivative of order $1/2\,. $
\vsp
Usually, the starting point to introduce the Riemann-Liouville fractional
calculus is the well-known Cauchy's iterated  formula, which
provides the  $n$-fold primitive of a given function $f(t)$
in terms of a single integral.
If $t > c \in \RR\,,$ it reads
$$ \eqalign{I^n_c \, f(t)&=
 \int_c^t \int_c^{\tau_{n-1}} \dots \int_c^{\tau_2}  \int_c^{\tau_1}
  \!\! f(\tau ) \, d\tau \, d\tau _1 \dots  d\tau _{n-1} \cr
  {} &=
  \rec{(n-1)!}\, \int_c^t \!\!	(t-\tau )^{n-1}\,f(\tau )\, d\tau \,,
  \q n=1,\, 2, \, \dots \,. \cr} \eqno(B.1)  $$
Here   $c$ denotes the point where  the primitive  is required to
vanish along with its first $n-1$ derivatives.
The passage from $n\in	\NN$ to $\alpha \in \RR^+$ is now quite natural
taking into account that $(n-1)! = \Gamma(n)\,. $
Consequently we define
\vsn --
\underbar{{\it the  fractional integral of $f(t)$ of order $\alpha $}}
 (with starting point $c$)
   $$ I_c^\alpha \, f(t):=
       \rec{\Gamma(\alpha )}\, \int_c^t \!\!  (t-\tau )^{\alpha -1}\,
    f(\tau )\, d\tau \,,   \q \alpha \in \RR^+\,. \eqno(B.2)$$
For $\alpha =0$ we define $  I_c^0 \, f(t) = f(t)$ so that
$I_c^0 \equiv \II$  where $\II$ is the identity operator.
The choice with $c=-\infty$ is originally due to Liouville (1832),
 while	with $c=0$ to Riemann (1847).
\vsp
In order to introduce the notion of {fractional derivative}
of order $\alpha \,,$
we need to consider the possibility to change
$\alpha \to -\alpha $ in  the r.h.s. of (B.2).
 While the extension in (B.1) from
$n$ to $\alpha > 0$  is quite legitimate,
the actual proposal requires
some care due to the convergence of the integral.
\vsp
If $\alpha $ denotes any positive real number
in the range $n-1 <\alpha < n$ with $n \in \NN$ and $f(t)$
is a sufficiently well-behaved function,
one usually  defines
\vsn  --
\underbar{{\it the  fractional derivative of $f(t)$
of order $\alpha $}} (with starting point $c$)
 $$
D_c^\alpha \,f(t) := {d^n \over dt^n} \, I_c^{n-\alpha } \,f(t) =
  \rec{\Gamma(n-\alpha )} \,{d^n  \over dt^n } \, \int_c^t
     {f(\tau )\over (t-\tau )^{\alpha +1- n}} \, d\tau
 \,.  \eqno(B.3)$$
\par
A  different formula  for the fractional derivative,
alternative to (B.3), originally introduced by Caputo [14-15], is
$$ \tilde D_c^\alpha \,f(t) :=	I_c^{n-\alpha }\,{d^n \over dt^n} \,f(t) =
  \rec{\Gamma(n-\alpha )} \, \, \int_c^t
     {f^{(n)}(\tau ) \over (t-\tau )^{\alpha +1- n}} \, d\tau
 \,.  \eqno(B.4)$$
 We  note  in general  that
$$  {d^n  \over dt^n } \, I_c^{n-\alpha} \, f(t)
 \ne I_c^{n-\alpha } \, {d^n  \over dt^n } \,f(t)\,,\eqno(B.5)$$
 unless   the function	$f(t)$ along with its first $n-1$ derivatives
  vanishes at $t=c^+$.
\vsp
For {\it causal} functions (\ie vanishing for $t<0\,$)
the choice $c=0\, $ is in order.
In this case it is convenient to introduce the so-called Gel'fand-Shilov
distribution [8]
$$ \Phi_\lambda(t) :=
   {t^{\lambda-1} \over \Gamma(\lambda )}\,  \Theta(t)
\,,    \qq \lambda  \in \CC \,,  \eqno(B.6)$$
where $ \, \Theta(t)\,$  is the unit step  Heaviside function
and $\Gamma(\lambda)$  is the Gamma function.
For $\lambda =-n \;(n=0\,,\,1\,,\,\ldots)\,,$
 $\Phi_\lambda (t)$ reduces to the $n$-derivative (in the generalized
sense) of  the {\it Dirac delta distribution},
$$    \Phi_{-n} (t) := {\;t^{-n-1} \over \Gamma(-n)} \, \Theta(t) =
  \delta ^{(n)}(t)\,, \q n=0\,,\,1\,,\ldots  \eqno(B.7)$$
Assuming that
the passage of the $n$-derivative in (B.3) under
integral is legitimate, one	
recognizes that,  for $ n-1 <\alpha <n \,,$
$$
    D_0^\alpha \, f(t) =
 \tilde D_0^\alpha \, f(t) +
  \sum_{k=0}^{n-1} f^{(k)}(0^+) \, \Phi_{(k-\alpha +1)} (t)\,,
   \eqno(B.8)$$
and,  using the (generalized) technique of
Laplace transforms,
$$	 {\cal{L}}\, \l\{  \tilde D_0^\alpha \,f(t) \r\} =
	s^\alpha\,  \bar f(s)
   -\sum_{k=0}^{n-1} s^{\alpha -1-k} \, f^{(k)}(0^+)\,.
 \eqno(B.9)$$
\vsp
Let us now consider the {\it causal} restriction of the Basset force.
We easily recognize in (2.6) that
$$  B(t)=	\rec{\Gamma (1/2)} \,
     \int_{0^-}^t {d V(\tau )/d\tau \over \sqrt{t-\tau}}\,d\tau
  =	   \rec{\Gamma (1/2)} \,
     \int_{0}^t {d V(\tau )/d\tau \over \sqrt{t-\tau}}\,d\tau
   + V(0) \, \Phi_{1/2} (t)\,. \eqno(B.10)    $$
Consequently, we can write the following equivalent expressions
in terms of the derivatives of order $1/2\,,$
$$ B(t) = \tilde D_0 ^{1/2} \, V(t) +	V(0) \, \Phi_{1/2} (t)
   = D_0^{1/2} \, V(t) \,. \eqno(B.11)$$
Applying the property (B.9) to (B.11), we can also write
$$ \bar B(s) = s^{1/2} \, \bar V(s) \, \Longleftrightarrow \,
  B(t) = \Phi_{-1/2}(t) \,*\, V(t)\,, \eqno(B.12)$$
where the convolution is to be intended in the generalized sense
of  Gel'fand-Shilov [8].
\vfill\eject

\def\rf#1{\item{$[{#1}]$ \ }}

\pageno=153
\cen{{REFERENCES}}\vs\vsp
\rf1  CHANDRASEKHAR, S., Stochastic problems in Physics and Astronomy,
 {\it Rev. Mod. Phys.}, {\bf 15} (1943), 1-89.
\vsn
\rf2 KUBO, R., The fluctuation -- dissipation theorem,
 {\it Report on Progress in Physics}, {\bf 29} Part I (1966), 255-284.
\vsn
\rf3 KUBO, R., TODA, M. and HASHITSUME, N.,
 {\it Statistical Physics II}, 2nd ed.,
  Springer Verlag, Berlin (1991).
\vsn
\rf4 MURALIDHAR, R., RAMKRISHNA, D., NAKANISHI, H. and JACOBS, D.,
 Anomalous diffusion: a dynamic perspective,
 {\it Physica}, {\bf A 167} (1990), 539-559.
\vsn
\rf5 WIDOM, A., Velocity fluctuations of a hard-core Brownian particle,
 {\it Phys. Rev.}, {\bf A 3}  (1971), 1394-1396.
\vsn
\rf6 CASE, K.M., Velocity fluctuations of a body in a fluid,
 {\it Phys. Fluids}, {\bf 14}  (1971), 2091-2095.
\vsn
\rf7 LANDAU, L.D. and LIFSHITZ, E.M.,
 {\it Fluid Mechanics},   
 Pergamon Press, Oxford (1987).
\vsn
\rf8  GEL'FAND, I.M.  and  SHILOV, G.E.,
  {\it Generalized Functions}, Vol. 1,
  Academic Press, New York (1964).
\vsn
\rf{9} BOUSSINESQ, J.,
 Sur la r\'esistance qu'oppose un liquid ind\'efini en repos,
 san pesanteur, au mouvement vari\'e d'une sph\`ere solide qu'il
 mouille sur toute sa surface, quand les vitesses restent bien
 continues et assez faibles pour que leurs carr\'es et
 produits soient n\'egligeables,
 {\it C.R. Acad. Paris}, {\bf 100} (1885), 935-937.
\vsn
\rf{10}  BASSET, A.B.,
  {\it A Treatise on Hydrodynamics}, Vol. 2 (Chap. 22,	pp. 285-297),
  Deighton Bell, Cambridge (1888).
\vsn
\rf{11} ALDER, B. J. and WAINWRIGHT, T.E.,
  Decay of velocity autocorrelation function,
  {\it Phys. Rev.}, {\bf A 1}  (1970), 18-21.
\vsn
\rf{12}  MAINARDI, F., PIRONI, P. and TAMPIERI, F.,
 On a generalization of the Basset problem via Fractional Calculus,
 in   B.  Tabarrok and S. Dost (Editors),
  {\it Proceedings CANCAM 95}, Vol. 2, pp. 836-837 (1995).
 [15-th Canadian Congress of Applied Mechanics,
    University of Victoria, B.C., Canada, 28 May - 2 June 1995]
\vsn
\rf{13}  MAINARDI, F., PIRONI, P. and TAMPIERI, F.,
  A numerical approach to the generalized Basset problem for a sphere
   accelerating in a viscous fluid,
   in	P.A.  Thibault and D.M. Bergeron (Editors),
   {\it Proceedings  CFD 95}, Vol. II, pp. 105-112 (1995).
   [3-rd Annual Conference of the Computational Fluid Dynamics Society
    of Canada, Banff, Alberta, Canada, 25-27 June 1995].
 \vsn
\rf{14}  MAINARDI, F.,
  Fractional calculus, some basic problems in
  continuum and statistical mechanics,
  in A. Carpinteri and F. Mainardi (Editors),
 {\it Fractals and Fractional Calculus in  Continuum Mechanics},
  Springer Verlag, Wien  (1997), pp. 291-348.
\vsn
\rf{15} MAINARDI, F. and TAMPIERI, F.,
 Diffusion regimes in Brownian motion induced by the Basset history
 force,
 Technical Paper No 8 (FISBAT-TP-98/1), FISBAT-CNR, Bologna 22 June 1998,
 pp. 20.
 \vsn
\rf{16}   CAPUTO, M.,  {\it Elasticit\`a e Dissipazione},
     Zanichelli, Bologna (1969) [In Italian].
\vsn
\rf{17} OLDHAM, K.B. and  SPANIER, J.,
 {\it The Fractional Calculus},
   Academic Press, New	York (1974).
\vsn
\rf{18}  SAMKO, S.G., KILBAS, A.A. and MARICHEV, O.I.,
 {\it Fractional   Integrals and Derivatives: Theory and Applications},
  Gordon and Breach, New York (1993)
 [Translated from the Russian edition (1987)].
\vsn
\rf{19}  MILLER, K.S. and   ROSS, B.,
 {\it An Introduction to the Fractional   Calculus and Fractional
Differential   Equations},   Wiley, New York (1993).
\vsn
\rf{20} GORENFLO, R. and MAINARDI, F.,
 Fractional calculus, integral and  differential equations of
   fractional order,
 in A. Carpinteri and F. Mainardi (Editors),
 {\it Fractals and Fractional Calculus in  Continuum Mechanics},
 Springer Verlag, Wien (1997), pp. 223-276.
\vsn
\bye